\documentclass[twoside,12pt]{article}
\usepackage{epsfig}
\usepackage{color} %only for highlighting changes

\newcommand{\be}{\begin{equation}}
\newcommand{\ee}{\end{equation}}
\newcommand{\bea}{\begin{eqnarray}}
\newcommand{\eea}{\end{eqnarray}}

\begin{document}

\title{ \vspace{1cm} Neutrinos in dense quark matter and cooling of compact 
stars}

\author{D.~Blaschke$^{1,2}$ and J.~Berdermann$^3$ \\
\\
$^1$Institute for Theoretical Physics, University of Wroclaw, Poland\\
$^2$Bogoliubov Laboratory for Theoretical Physics, JINR Dubna, Russia\\
$^3$DESY, Platanenallee 6, D-15738 Zeuthen, Germany }
\maketitle
\begin{abstract} 
We discuss that observational constraints on neutrino cooling processes may 
restrict the spectrum of quark matter phases admissible for compact star 
interiors.
%We discuss aspects to be taken into consideration when discussing the neutrino
%cooling processes of compact stars with deconfined quark matter in their 
%interiors.
\end{abstract}
%\eject
%\tableofcontents
\section{Introduction}
The existence of quark matter in compact star (CS) interiors is debated 
controversially since observations of the thermal emission from objects like 
RXJ 1856.5-3754 \cite{Trumper:2003we} indicate large masses and/or 
luminosity radii requiring a stiff equation of state (EoS) at high densities.
Although this seemingly excludes a phase transition accompanied with a 
softening of the EoS, it has been demonstrated \cite{Alford:2006vz} that 
modern QCD motivated theories could fulfill such observational constraints, 
e.g., due to the presence of vector mean fields which stiffen the quark matter 
EoS  \cite{Klahn:2006iw}. 
Investigations of color superconducting quark matter phases (for a recent 
review, see \cite{Alford:2007xm}) show that large diquark pairing gaps of 
$\sim 10 - 100$ MeV are possible, which lower the critical density for the 
chiral symmetry restoration due to their competition with the 
chiral condensate. This entails early quark deconfinement unless color forces 
remain strong enough to confine even (almost) massless quarks in a new phase 
of chirally symmetric hadronic matter, the hypothetic quarkyonic phase 
\cite{McLerran:2007qj,Andronic:2009gj}. 
Here, we will not yet consider quarkyonic matter, but discuss the viability of 
three-flavor (CFL), 
two-flavor (2SC) and one-flavor (d-CSL) color superconducting phases under 
constraints for mass and radius as well as CS cooling.

Spin-0 phases which pair quarks of different flavor and opposite spins with
large gaps (2SC and CFL) might be too fragile to withstand the stress of
flavor asymmetry and strong magnetic fields in neutron stars. 
It is not clarified yet whether the occurrence of the celebrated CFL phase 
in the core of a hybrid star leads necessarily to a gravitational instability 
\cite{Klahn:2006iw} as a result
of the strong softening of the EoS (for an exception, see 
\cite{Pagliara:2007ph}).
Disregarding the hadronic shell even interesting mass twin sequences are 
possible \cite{Sandin:2007zr}. 
If hybrid stars with CFL quark matter cores may form a third CS family, this 
might allow for a new paradigm to explain the Janus-faced CS phenomenology:
large luminosity radii vs. millisecond rotation periods, fast cooling vs. 
slow cooling etc.

After the new color superconductivity phases with large Spin-0 pairing gaps 
(2SC and CFL) were suggested in 1997, a discussion of late time cooling 
scenarios has been performed in \cite{Blaschke:1999qx,Page:2000wt}, 
followed by a full transport calculation with nontrivial temperature profile
evolution in hybrid stars with CSC quark matter cores \cite{Blaschke:2000dy}.
These calculations showed that the pairing pattern where all quarks are paired 
with large gaps delay the cooling and may be excluded by the observation of 
fast coolers like the Vela pulsar, unless they form a third family.

In these works as well as in the detailed study in Ref.~\cite{Jaikumar:2005hy}
the fact was ignored that in the 2SC phase the quarks with one of the colors 
(e.g., the blue ones) remain unpaired so that the fast direct Urca (DU) process
is operative, entailing too fast cooling. 
For heuristic purposes, a residual single-color pairing of the blue quarks 
(X-gap) has been introduced \cite{Grigorian:2004jq} and it has been found that 
the best cooling phenomenology would be obtained for hybrid 
stars with superconducting quark matter cores in the 2SC+X phase when 
this smallest gap is in the range between 100 keV and 1 MeV, with a decreasing
density dependence \cite{Blaschke:2005dc,Popov:2005xa}.
An alternative to the postulated 2SC+X phase which provides a microscopic 
approach to the pairing pattern in accordance with cooling phenomenology is 
the isotropic color-spin-locking (CSL) phase \cite{Aguilera:2005tg} for which 
quarks of the same flavor are paired in a spin-1 state. This phase is thus 
rather inert against the neutron star stress. 

Recently, it has been argued that the different quark flavors could occur 
sequentially, i.e. at different threshold densities in neutron star matter
\cite{Blaschke:2008br,Blaschke:2008vh}. In analogy to the 
neutron drip in the crust, a down-quark drip density can occur in the core,
when d-quarks undergo a chiral restoration and partial nucleon dissociation 
leads to the formation of a superconducting  single-flavor subphase (d-CSL) 
immersed in nuclear matter.
This mixture of d-CSL and nuclear matter can extend from the interior to the
crust core boundary in stable hybrid star configurations which fulfill the 
stringent mass-radius constraints mentioned above.
It also bears interesting perspectives for the neutrino transport and cooling
properties \cite{Blaschke:2008br,Blaschke:2007bv} which we 
discuss now a bit more in detail. 
\section{Neutrino emissivity and deep crustal heating}    

According to \cite{Stejner:2006tj}, two interesting phenomena, the
superbursters and the soft X-ray transients may be probe the existence of 
CS deep crustal heating processes subject to the following constraints\\
1. A thin baryonic crust of the star with a width between 100 to 400 m.\\
2. An energy release of 1 to 100 MeV per accreted nucleon by 
conversion at the crust-core boundary.\\
3. The thermal conductivity %, $\kappa$, of the underlying quark matter phase
 has to be in the range of $10^{19}-10^{22}~{\rm erg~cm^{-1}~s^{-1}~K^{-1}}$.\\
4. The fast DU neutrino emissivity process should be strongly
 suppressed or not operational.\\ 
Following the suggestion of Refs.~\cite{Page:2005ky,Cumming:2005kk},
the conversion of nuclear matter to CFL strange quark matter, powered by 
continuous accretion in a LMXB may provide a mechanism for the deep crustal 
heating. 
As the nuclei penetrate into the quark matter core a conversion energy
between 1 to 100 MeV is released per accreted nucleon. 
This heats the core until an equilibrium temperature is reached 
and the heating is balanced by neutrino emission.
As the DU neutrino emission process is strongly suppressed by large 
CFL gaps, the heat produced at the crust-core boundary is not radiated away 
but conducted so that the fusion ignition condition gets fulfilled.
\\
In \cite{Blaschke:2008br} we have shown that a d-quark CSL/nuclear phase 
mixture can also fulfill the above constraints on the deep crustal heating 
mechanism. In particular, we estimated the heat per accreted nucleon from the 
d-quark drip due to partial chiral restoration to the order of $10$ MeV and 
the DU process which we exclude in the nuclear subphase does also not 
occur in d-CSL subphase since the Fermi sea of up quarks is closed. 
An interesting question for the d-CSL/nuclear matter phase to be clarified is
the possible role of confining interactions between the colored d-quarks which 
would energetically forbid a rather dilute d-quark admixture and sets a 
threshold for the d-quark drip. It may well be that the proper account for 
confining interactions within a chiral model theory would lead to a 
modification of the nucleon properties rather than to their partial 
dissociation at the chiral restoration for the down quarks. 
A modified quarkyonic phase for isospin asymmetric matter could be suggested.

\section{Perspectives}
Hybrid Stars with a mixture of nuclear matter and d-CSL quark matter phases 
fulfill not only stringent constraints for large radii and masses but provide
also a viable deep crustal heating process. This mixed phase should probably 
better described as a kind of quarkyonic matter, once this gets accessible 
to a theoretical description.
Further investigation needs the idea that CFL quark core stars form a third 
CS family providing a possible explanation to a different class of CS  
phenomena as, e.g., GRBs \cite{Ouyed:2005tm,Berdermann:2006rk}, 
SGRs, AXPs and XDINs \cite{Niebergal:2009yb}.

\subsection*{Acknowledgements}
We are grateful to our colleagues 
%, in particular to D.N.~Aguilera, M.~Buballa,
%H.~Grigorian, T.~Kl\"ahn, S.~Popov, F.~Sandin, D.N.~Voskresensky and F.~Weber
for discussions and collaboration.
D.B. acknowledges an EPS Fellowship, %from the European Physical Society 
%for the participation in the School 
and partial support by MNiSW
%Ministry for Science and Higher Education under 
grant No. N N 202 231837, by RFBR
%the Russian Fund for Fundamental Investigations under 
grant No. 08-02-01003-a and 
by CompStar, a Research Networking Programme of the European Science 
Foundation.

\end{document}